\documentclass[11pt]{article}
\usepackage{latexsym,epsfig,a4wide,multirow,multicol}

\title{Comparison of Pedestrian Fundamental Diagram Across Cultures}

\author{Ujjal Chattaraj$^{1,2}$, Armin Seyfried$^1$, Partha Chakroborty$^2$ \\
[0.2cm]
\footnotesize $^1$J\"ulich Supercomputing Centre, Research Centre J\"ulich,
52425 J\"ulich, Germany\\
[-0.1cm]\footnotesize $^2$Indian Institute of Technology Kanpur, Kanpur 208
016, India\\
\footnotesize  E-mail: ujjal@iitk.ac.in, a.seyfried@fz-juelich.de, partha@iitk.ac.in\\
}

\begin{document}

\maketitle

\begin{abstract}

\par
\noindent


The relation between speed and density is connected with every self-organization 
phenomenon of pedestrian dynamics and offers the opportunity to analyze them 
quantitatively. But even for the simplest systems, like pedestrian streams in 
corridors, this fundamental relation isn't completely understood. Specifications 
of this characteristic in guidelines and text books differ considerably 
reflecting the contradictory database and the controversial discussion 
documented in the literature. In this contribution it is studied whether 
cultural influences and length of the corridor can be the causes for these 
deviations. To reduce as much as possible unintentioned effects, a system is 
chosen with reduced degrees of freedom and thus the most simple system, namely 
the movement of pedestrians along a line under closed boundary conditions. It 
is found that the speed of Indian test persons is less dependent on density 
than the speed of German test persons. Surprisingly the more unordered behaviour 
of the Indians is more effective than the ordered behaviour of the Germans. 
Without any statistical measure one cannot conclude about whether there are 
differences or not. By hypothesis test it is found quantitatively that these 
differences exist, suggesting cultural differences in the fundamental diagram 
of pedestrians.\\



\end{abstract}



\section{Introduction}
\label{INTRO}

Dimension of pedestrian facilities is relevant in respect of comfort and
safety in buildings with a large number of occupants (or people).
Concerning comfort as well as safety several planning guidelines and text books
are available
\cite{DiNenno2002,Oeding1963,Fruin1971,Weidmann1993,Nelson2002,Predtechenskii1978,HCM,Kholshevnikov2008}.
In all these books or guidelines one basic characteristic to describe
the dynamics of pedestrians is the relation between density and flow or density
and speed, which following the terminology of vehicular traffic is also named
the fundamental diagram. To illustrate the importance of the fundamental diagram
two frequently asked questions that appear while designing of pedestrian
facilities is discussed. In respect of safety it is important to know that how
many pedestrians can pass a certain cross-section of the facility. Thus one 
wants to know how many pedestrians $\Delta N$ can pass the facility in a given 
time interval $\Delta t$.  The quotient $\Delta N / \Delta t$ gives the flow $J$. 
Obviously the flow depends on the density. Another example is the calculation 
of travel times. Then one wants to know how long it will take to reach a 
certain destination, or an exit. The describing quantity is the speed which 
again will depend on the density. For low density one will be able to walk with 
the free flow speed. However with increasing density the speed will decrease and 
thus the travel-time increase. Thus the  basic quantities to describe pedestrian
stream are, the speed $v$, the density $\rho$ and the flow $J$. These
quantities are related by the flow equation $J=\rho\,v\,w$ where $w$ gives the
width of the facility and the empirical relation between  flow and speed $J(v)$ 
or the relation between speed and density $v(\rho)$. For different
facilities like stairs and corridors the shape of the diagrams differ, but
in general it is assumed that the fundamental diagrams for the same type of
facilities but different widths merge into one diagram for the specific 
flow $J_s=\frac{J}{w}$.

\begin{figure}[htb]
\centering
   \includegraphics[width=0.48\columnwidth]{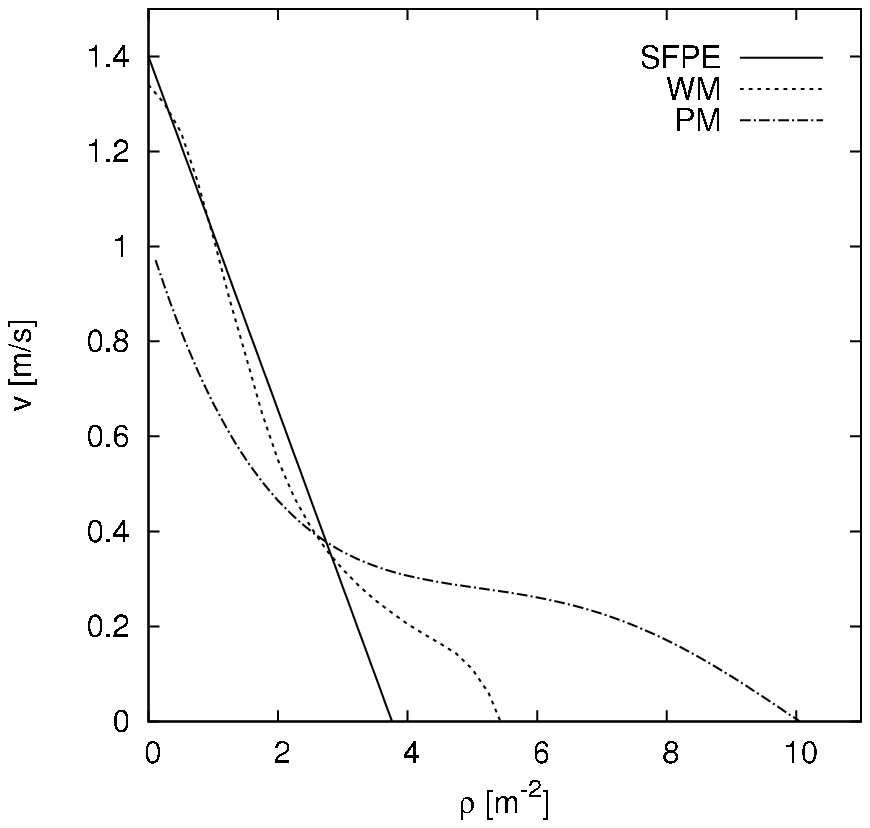}
\quad
   \includegraphics[width=0.48\columnwidth]{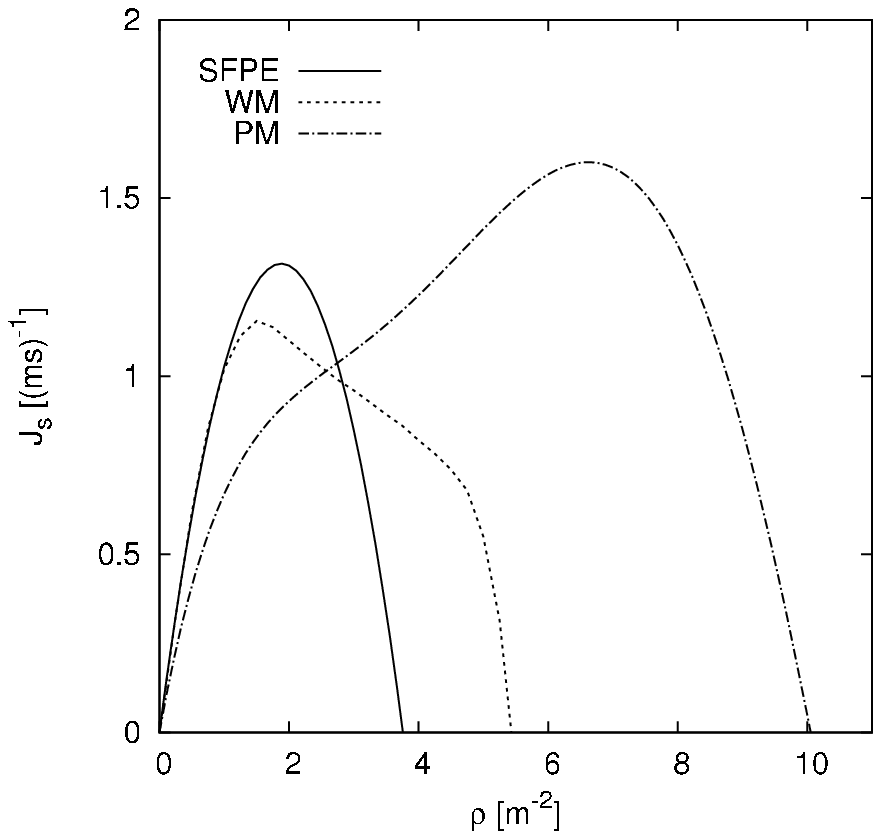}
\caption{Specifications for the fundamental diagram in different
  guidelines. (SFPE: SFPE Handbook of Fire Protection Engineering
  \cite{DiNenno2002}, PM: Planning for foot traffic flow in buildings
  \cite{Predtechenskii1978} and WM the guideline 'Transporttechnik der
  Fussg\"anger' \cite{Weidmann1993}.}
\label{FD-LIT}
\end{figure}

Figure \ref{FD-LIT} shows various fundamental diagrams used in planing
guidelines \cite{Predtechenskii1978,Weidmann1993,Nelson2002}. The comparison
reveals that specifications disagree considerably. In particular the maximum
of the function giving the capacity $J_{s,max}$ ranges from $1.2\;$(ms)$^{-1}$
to $1.8\;$(ms)$^{-1}$, the density value where the maximum flow is 
reached $\rho_c$ (optimal density) ranges from $1.75\;$m$^{-2}$ to $7\;$m$^{-2}$ 
and, most notably, the density $\rho_0$ (jam density) where the speed approaches
zero due to overcrowding ranges from $3.8\;$m$^{-2}$ to $10\;$m$^{-2}$. For a
detailed discussion of the causes for these deviation refer to
\cite{Schadschneider}.

Several explanations for these deviations have been suggested, including
cultural and population differences \cite{Morrall1991,Helbing2007},
differences between uni- and multi-directional flow
\cite{Navin1969,Pushkarev1975,Lam2003}, short-ranged fluctuations
\cite{Pushkarev1975}, influence of psychological factors given by the
incentive of the movement \cite{Predtechenskii1978, Kholshevnikov2008} and,
partially related to the latter, the type of traffic (commuters, shoppers)
\cite{Oeding1963}.

However there is no common agreement in the community which factors influence
the relation and which do not. For example Weidmann \cite{Weidmann1993}
neglects differences between uni- and multi-directional flow in accordance with 
Fruin, who states in his often cited book \cite{Fruin1971} that the fundamental
diagrams of multi-directional and unidirectional flow differ only slightly.
This disagrees with results of Navin and Wheeler \cite{Navin1969} and
Lam et al.\ \cite{Lam2003} who found a reduction of the flow due to
directional imbalances. Unfortunately most of the authors do not give all
informations and a comparison of different studies in general is subjected to 
uncontrollable errors and uncertainties. In addition it was shown in \cite{Seyfried2008b}
that the measurement methods influence the resulting relations, restricting
the number of comparable studies.

However, all diagrams agree in one characteristic: speed decreases
with increasing density. So the discussion above indicates
there are many possible reasons and causes for the speed reduction.
For the movement of pedestrians along a line a linear relation between
speed and the inverse of the density was measured in \cite{Seyfried2005}.
The speed for walking pedestrians depends also linearly on the length of a
stride (or step size) \cite{Weidmann1993} and the inverse of the density can be 
regarded as the required length of one pedestrian to move. Thus it seems that 
smaller step sizes caused by a reduction of the available space with
increasing density is, at least for a certain density region, one cause
for the decrease of speed.

The discussion above shows that there are many possible factors which influence
the fundamental diagram. To identify the factors that influence the fundamental 
diagram it is necessary to exclude as much as possible, influences of measurement 
methodology and short range fluctuations on the data. Moreover the study 
presented in \cite{Seyfried2005} showed that a reduction of the degrees of 
freedom gives an idea about the causes responsible
for the speed reduction with density in pedestrian traffic.

In line with this approach experiments under laboratory
conditions are performed and a system is chosen with smaller degrees of freedom
to study whether and how cultural differences influences the speed density
relation. The study of {\tt http://www.paceoflife.co.uk} and Morral
et.~al.~\cite{Morrall1991} is restricted to low densities 
$\rho<1\;$m$^{-2}$ only. Thus up to now it is not sure whether cultural
differences are present for the high density regime of the fundamental
diagram. In addition whether length of corridor impacts the results or not is
not studied yet. 

The problem at hand is to study pedestrian speed-distance headway data from
Germany and India. Further, the intention is to analyze parameters such as
free flow speed, minimum personal space and change of speed with distance
headway, and see whether these vary between cultures. For this purpose
experiments, as described in next section, were conducted in Germany and
India. The data from these experiments are analyzed for the various parameters
listed above and the analysis and results are presented later. The reason for
studying speed-distance headway relationship instead of speed-density
relationship is that it was intended to compare results from different
cultures by some quantitative measure. This can only be done easily if the
relationship is linear. It is a known fact and also was observed from the data
speed-distance headway relationship can be reasonably described through a
linear relation but speed-density can not be.


\section{Experimental Set-up and Data Collection}

\subsection{Experimental set-up}

The corridor is built up with chairs and ropes. The size and shape of the
corridor is same as adopted in \cite{Seyfried2005} for the similar experiment
in Germany. The length of the corridor is $l_p = 17.3$ m and that of the
measured section is $l_m = 2$ m. The measured section was constructed by
erecting two ranging rods at the entry and exit line of the measured section,
so that when a person is exactly at those positions the time can be noted. The
camera was set at a sufficient distance of $10$ m from the measured section
along the perpendicular bisector of the measured section to reduce parallax error.

\begin{figure}[htb]
\centering
    \includegraphics[width=0.44\columnwidth]{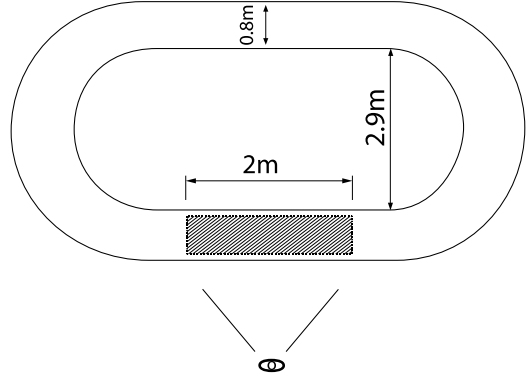}
\caption{Sketch of the experimental set-up in India and Germany.}
\label{SKETCH}
\end{figure}

The channel in which pedestrian motion is studied is shown in Figure \ref{SKETCH}. 
Data however is collected only for the shaded section indicated in the
figure. The width of the passageway in the straight section is $0.8$ m which is
sufficient for single file movement but not for overtaking. In the curved
section the width is increased to a maximum of $1.2$ m through elliptic
transition curves. The reason for this is that the curved portion of the flow
space may reduce speed. The experiment in India was done outdoor but on paved
ground. The subjects for the experiment consisted of graduate students,
technical staff of Indian Institute of Technology, Kanpur and local
residents of Kanpur city thus putting variation in population. They were
instructed not to overtake and not to push others. To obtain data on various
densities six sets of experiments with number of subjects $N = 1, 15, 20, 25,
30, 34$ were performed. For the experiments (except $N$ = $1$) all the subjects
used in that cycle were initially distributed uniformly in the corridor. After
the instruction to start was given every subject goes around the corridor two
times. After that an opening is created in the closed corridor through which
the subjects are allowed to leave and keep walking for a reasonable distance
away from the corridor so that there is no tailback effect.

\begin{figure}[htb]
\centering
    \includegraphics[width=0.44\columnwidth]{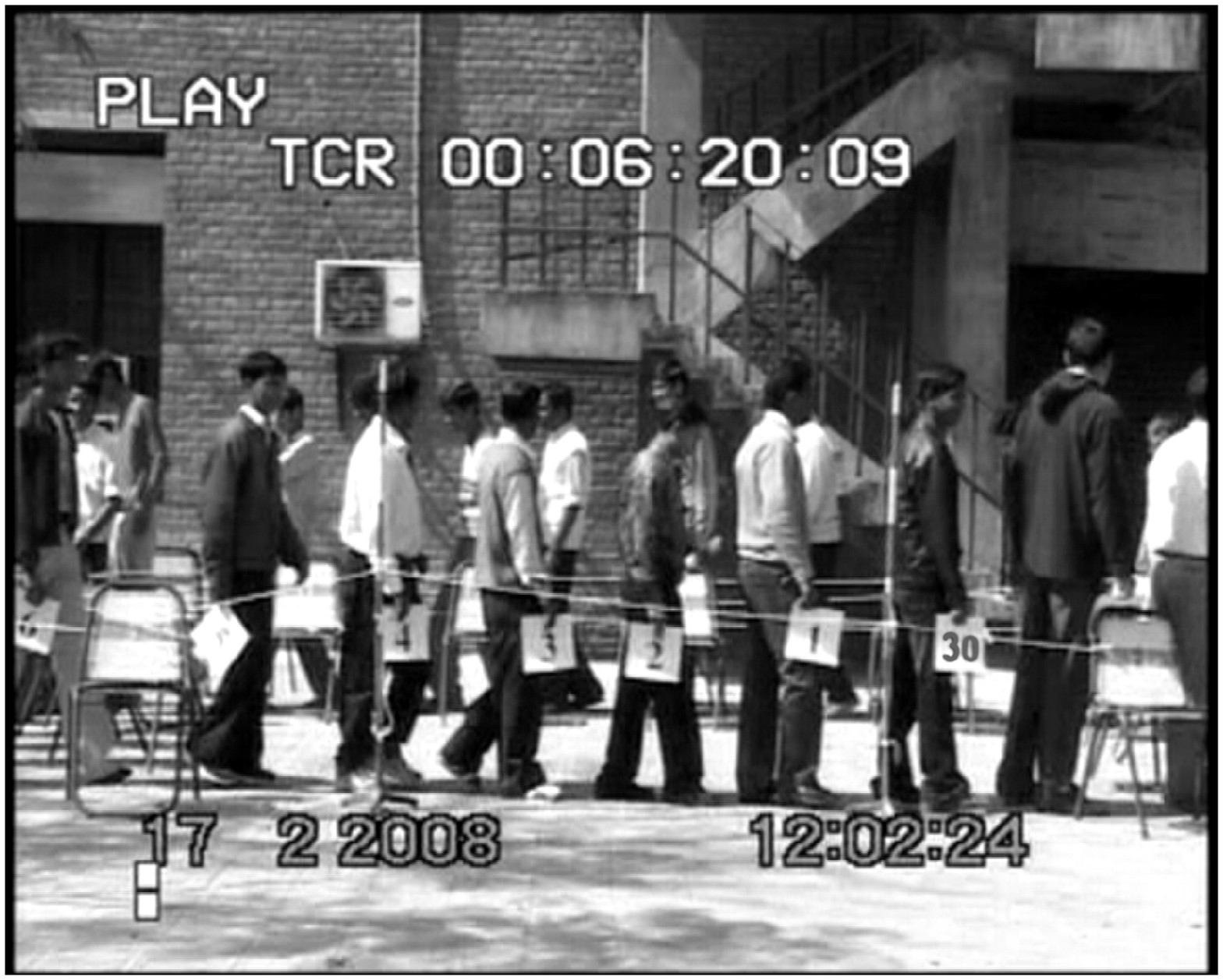}
\quad
\quad
    \includegraphics[width=0.455\columnwidth]{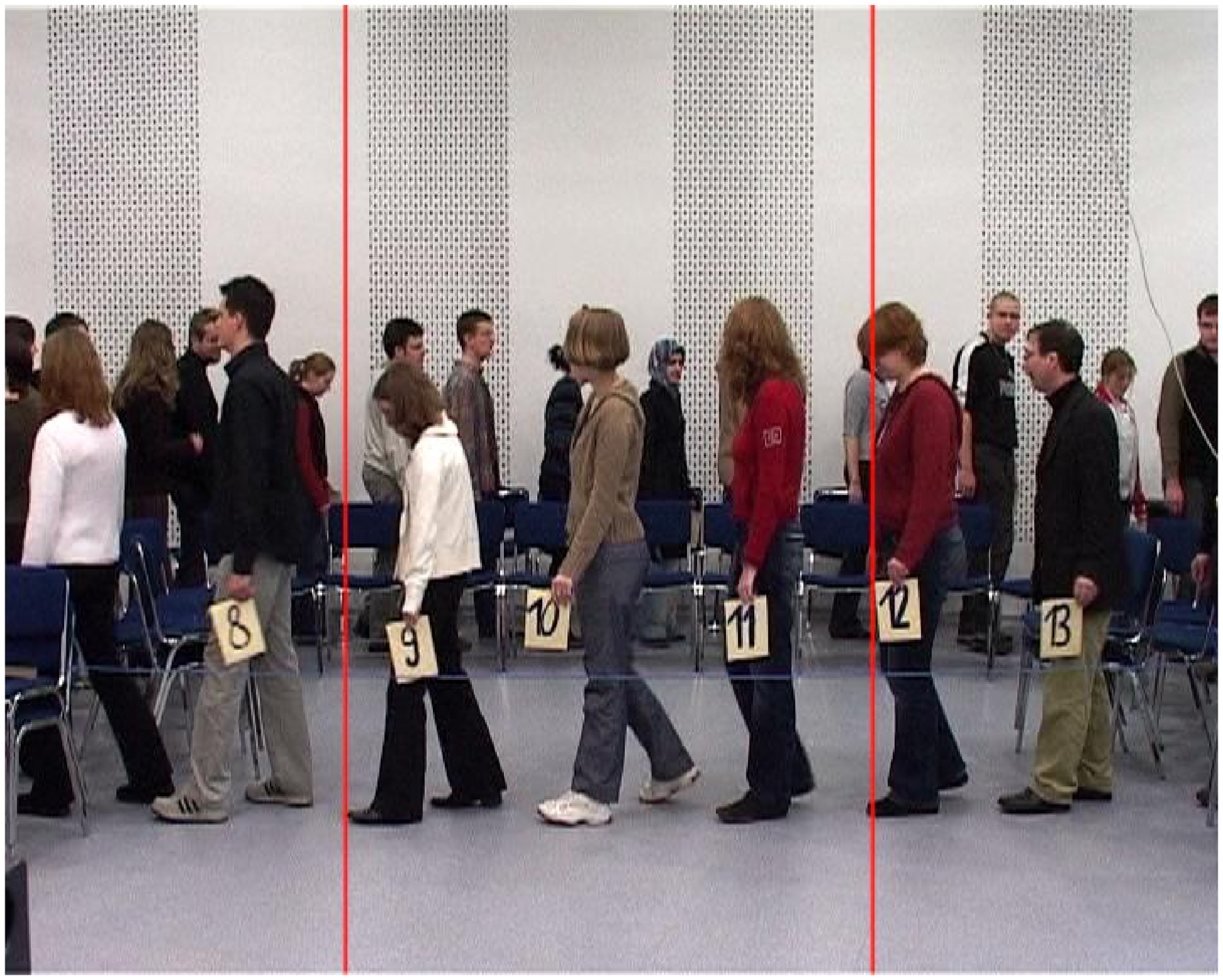}
\caption{Snapshots for the run with $N=30$ {\bf Left:} India  {\bf Right:}
Germany.
} \label{SNAPS}
\end{figure}

The experimental set-up for the German study is described in detail in
\cite{Seyfried2005}. The length and the width of the corridor, the position
and dimension of the measurement area, the number of subjects $N$ for the
six sets, the instruction of the test persons, as well as the measurement
method are identical to the study in India. Differences are the moving
direction and the composition of the test subjects. In the German study the
group of test persons was composed of male and female students and staff of
the Institute. In the Indian study the group consist of male
only. The moving direction of the experiment in India was clockwise while in
Germany it was anti-clockwise. Figure \ref{SNAPS} shows snapshots of the
Indian and German experiment.

\subsection{Data collection}

Initially speed-density data were collected and later density was converted to
distance headway by taking the reciprocal. To gather the speed-density data a
digital video camera ($25$ frames/second) was stationed as shown in Figure $2$.
In the Indian experiment ranging rods were used to demarcate the measured section
of Figure \ref{SKETCH}. A snapshot of the measured section obtained from video
data is shown in Figure \ref{SNAPS}. From the video data times at which an
individual (say individual $i$) entered ($t_i^{in}$) and left ($t_i^{out}$)
this section are noted for every individual. The times are noted to the 
nearest $0.04$ seconds. From this time points the information on speed and density are
obtained. The German data were collected in the same way, only the ranging rods
were not installed in the real set-up but lines added by image processing in
the video, see Figure~\ref{SNAPS}.


\section{Analysis and Results}

\subsection{Time development}

\begin{figure}[thb]
\centering
    \includegraphics[width=0.48\columnwidth,angle=0]{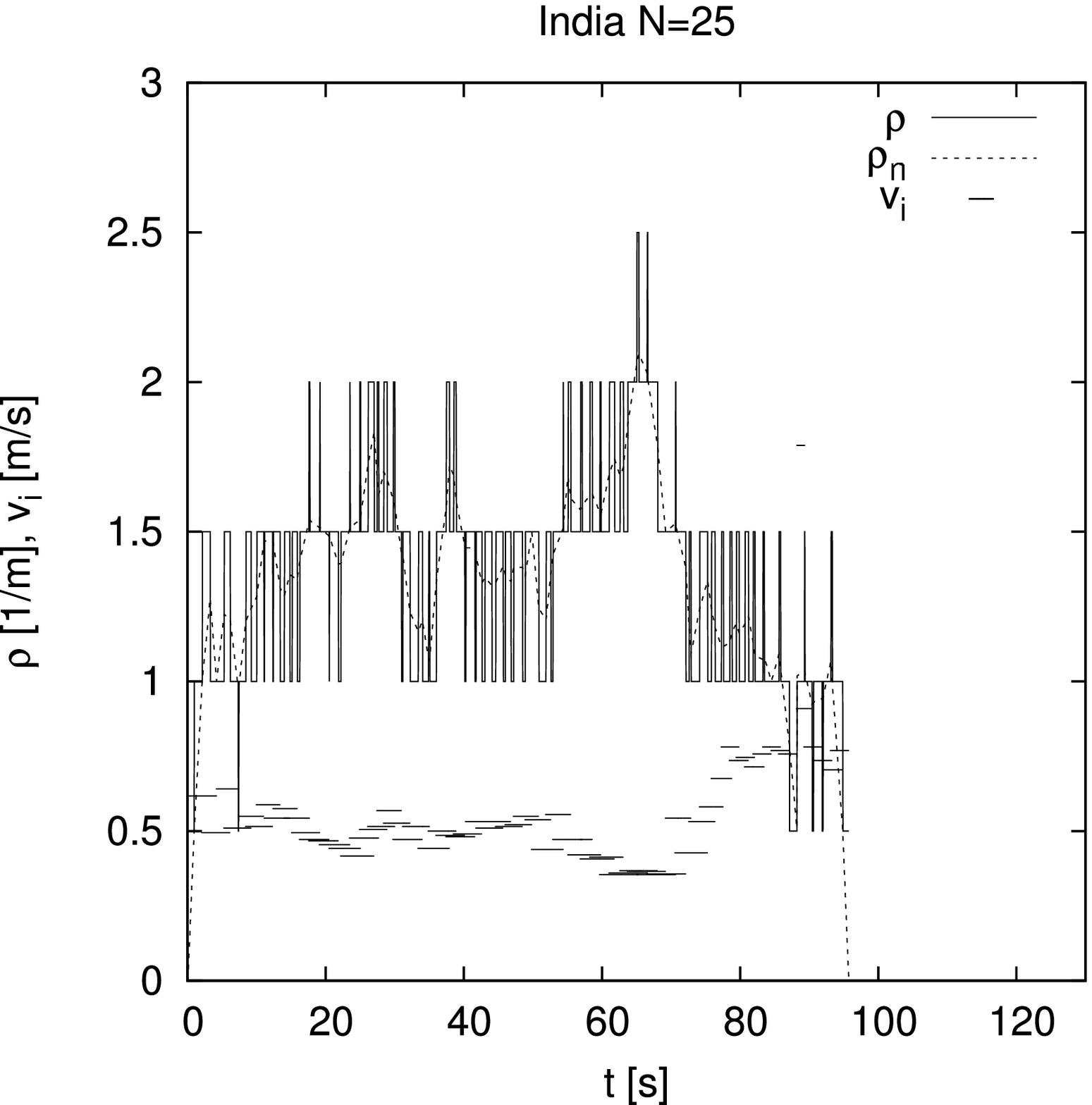}
\quad
    \includegraphics[width=0.48\columnwidth,angle=0]{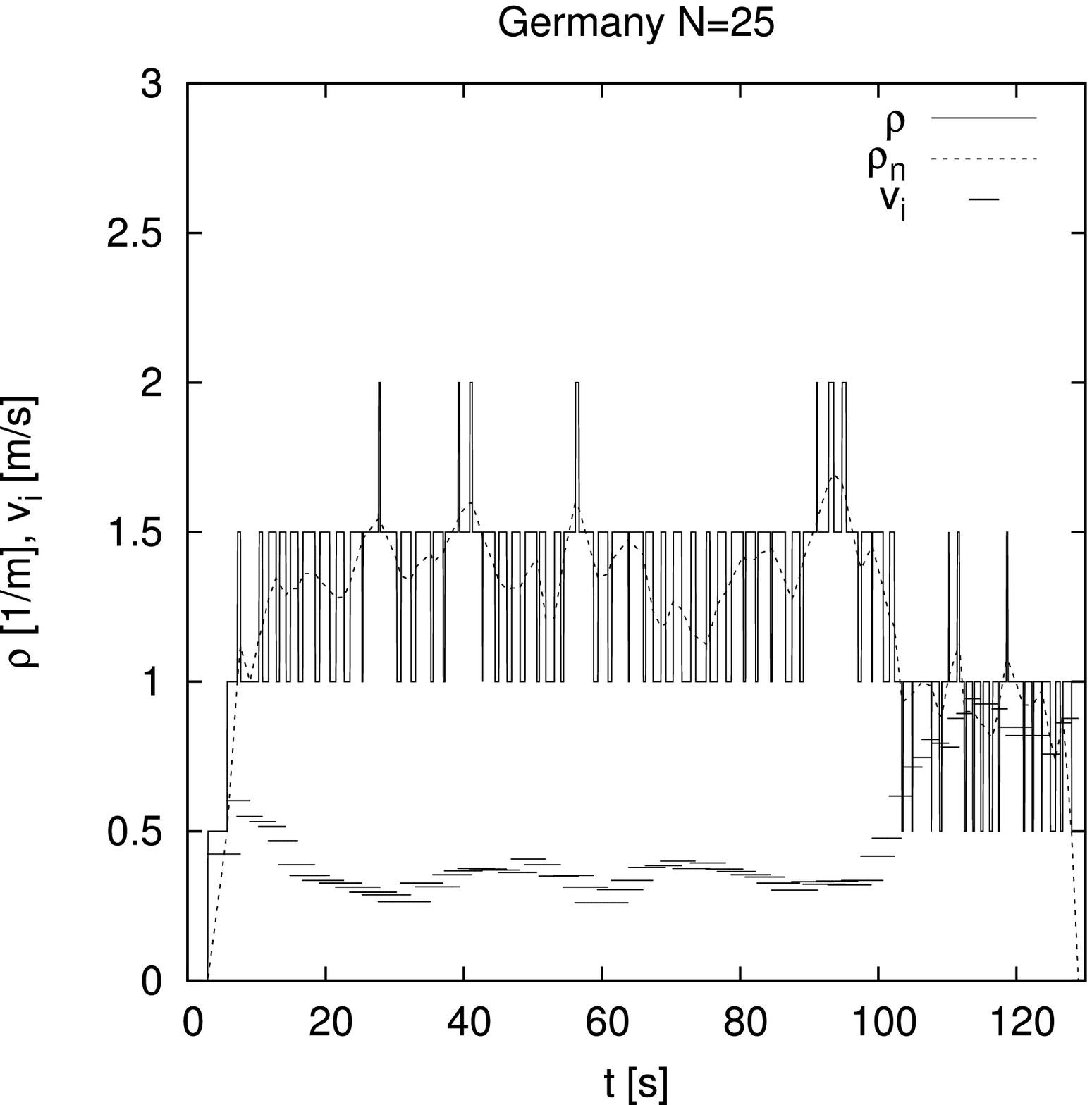}
\caption{Time development of the speed $v_i$ and density $\rho$ for the
  sets with $N=25$. The length of the lines $v_i$ indicate the time interval
  the pedestrian is inside of the measurement area. The classical density
  $\rho$ shows strong fluctuations while $\rho_n$, the density according the
  enhanced definition introduced in \cite{Seyfried2005}, is smooth. For the
  Indian ({\bf left}) and the German ({\bf right}) cases three phases can be
  identified: A starting phase, a steady state and end phase where the
  subjects leave the corridor.}
\label{TIMEDEV}
\end{figure}

Entrance and exit times $(t_i^{in},t_i^{out})$ were used to calculate the mean
speed of individual persons $v_i=\frac{l_m}{t_i^{out}-t_i^{in}}$ during the
crossing of the measurement section and the density $\rho(t)=N(t)/l_{m}$, see
Figure~\ref{TIMEDEV}. The classical density shows strong fluctuation and thus
the enhanced definition $\rho_n$ according to equation $1$ in
\cite{Seyfried2005} is used. The developing of the density $\rho_n$ and the
speed $v_i$ shows the distinct correlation of these quantities. For the Indian 
and the German experiment three phases can be identified: A starting phase, a
steady state and an end phase. In the starting phase the pedestrian begin to
walk and the value as well as the fluctuation of the speed differ from the
stationary state. During the stationary state the fluctuations in the Indian
data are higher than the fluctuations in the German data indicating a more
unordered character of movement in the Indian data. The end phase, indicated
by higher speeds starts once the corridor was opened and the persons leave the 
corridor. For the following analysis we restrict the data to the stationary state. 
The density assigned to the space mean speed of individual $i$, $v_i$ is the mean
value of the density during the crossing $\rho_i = \int dt \, \rho_n(t)$. It is 
noted that the differences between the enhanced and the classical density in the 
mean value over time are negligible.

\subsection{Study on free flow speed}

The linear distance headway - speed ($h$-$v$) plot of closed corridor can not
predict free flow speed. However data on free flow speed were collected by
making only one person move in the corridor. Mean free flow speed from Indian
data and German data are $1.27$ ms$^{-1}$ (with standard error of $0.03$) and
$1.24$ ms$^{-1}$ (with standard error of $0.02$), respectively. Hypothesis testing
on average free flow speed was conducted. Here lesser number of data points
were obtained, so a $t$-test is appropriate for hypothesis testing, assuming
normal distribution of data. The hypothesis that the mean free flow speed
obtained from India $(v_0^{I})$ and that from Germany $(v_0^{G})$ are the same
(i.e. Null hypothesis $H_0:\;v_0^{I} -v_0^{G}= 0$ and alternate hypothesis
$H_1:\;v_0^{I}-v_0^{G}\neq 0$) was tested. If the value of the expression 

\begin{equation}
  t = \frac{v_0^{I}-v_0^{G}}{\sqrt{S^2_{vI}+S^2_{vG}}},
  \label{TSTAT}
\end{equation}

comes out to be greater than some critical value at a certain level of
confidence the Null hypothesis can be rejected saying that there is difference
in free flow speed between Indian data and German data, otherwise not. 
Here, $S_{vI}$ and $S_{vG}$ are the standard errors for the free flow speed data
from India and Germany, respectively. The value of the above expression for
this data comes out to be $0.85$. The quantity obtained in equation \ref{TSTAT} follows 
a $t$-distribution with $df$ degrees of freedom. The value of $df$ is given by
Welch-Aspin equation as

\begin{equation}
df=\frac{(S_{vI}^2+S_{vG}^2)^2}{\frac{S_{vI}^4}{n_I-1}+\frac{S_{vG}^4}{n_G-1}}=33.
\end{equation}

Where, $n_I$ and $n_G$ are the number of data points for Indian and German
free flow speed data, respectively. For a two tailed $t$-test with a degree of
freedom $33$ the value of $t_{critical}$ is $2.03$ at $95\%$ level of
confidence. Since the value of $t$ is less than $t_{critical}$ the Null
hypothesis can not be rejected. The conclusion is that (mixed) population in
Germany and (male) population in India have free flow speeds, which are not
statistically different. This indicates that left to themselves both Germans
and Indians walk in a similar manner.

\subsection{Speed - density relation across cultures}

\begin{figure}[htb]
\centering
    \includegraphics[width=0.48\columnwidth,angle=-90]{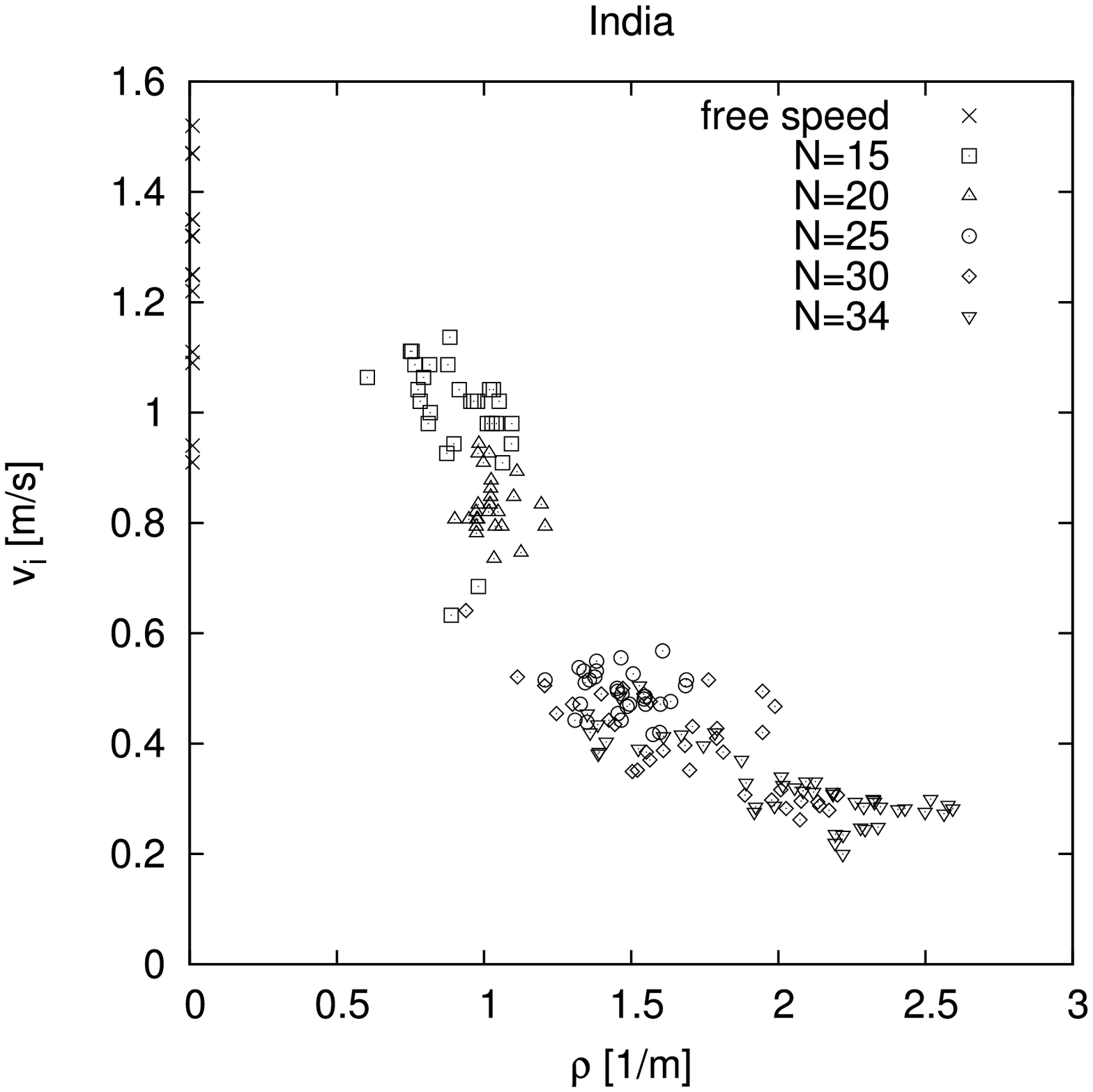}
\quad
    \includegraphics[width=0.48\columnwidth,angle=-90]{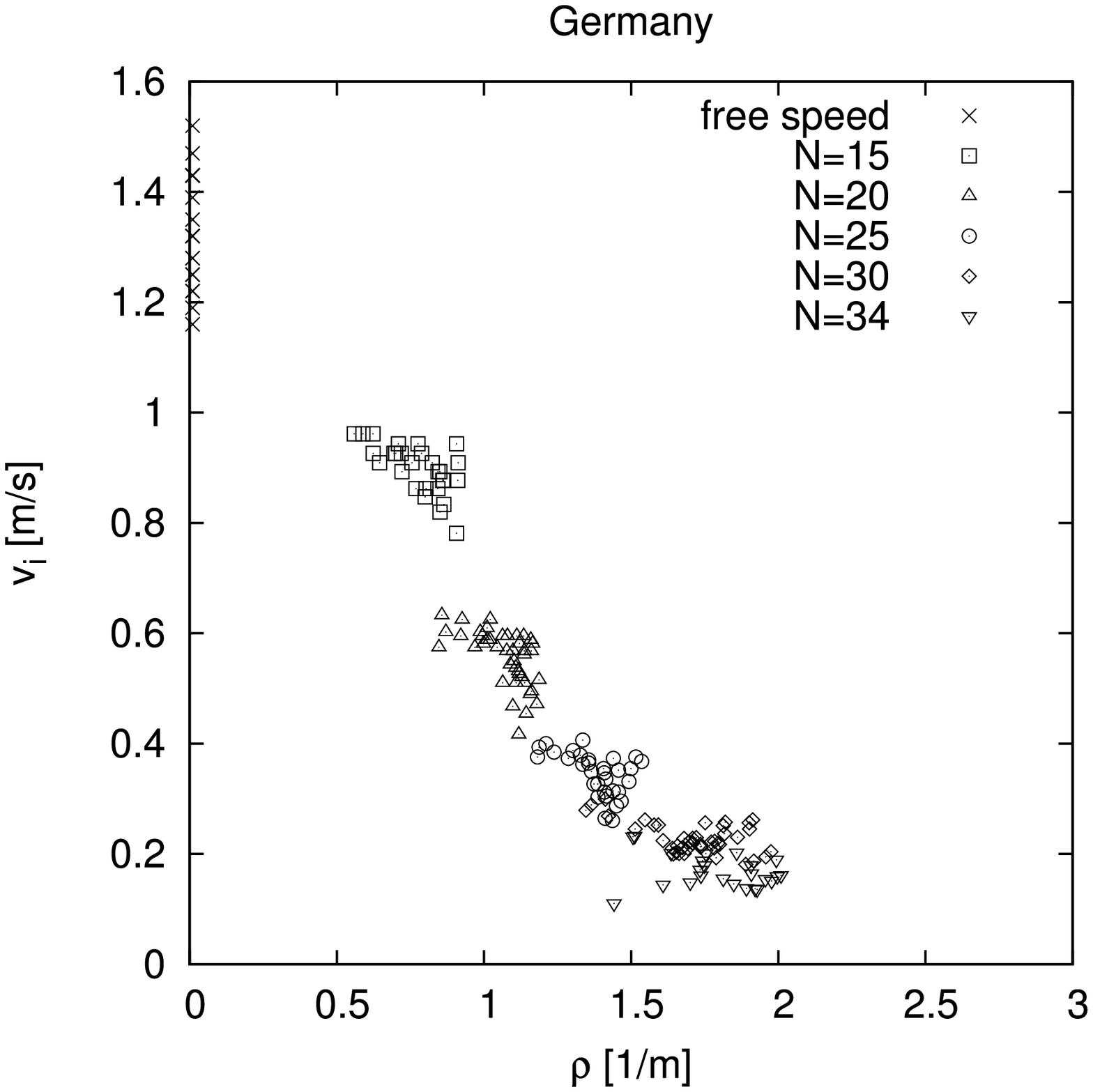}
\caption{Speed-density data from Indian (left) and German (right) study,
respectively.
} \label{RHO-V-ALL}
\end{figure}

The speed ($v$) - density ($\rho$) data in closed corridor condition obtained
from India and Germany are plotted in Figure \ref{RHO-V-ALL}.

On looking at the data the following subjective observations can be made.
First, for
every density region (except the free flow speed) Indians walk a bit
faster than Germans. This will be discussed in detail in the next section.
The free flow speed in India $v_0^{I}=1.27\,(\pm 0.16)$ ms$^{-1}$ is same as
the free flow speed in Germany $v_0^{G}=1.24\,(\pm 0.15)$ ms$^{-1}$. In spite of
averaging over time the scatter of Indian data is larger than in Germany.
e.g. for $N=15$ the speed ranges from $0.63$ ms$^{-1}$ to $1.11$ ms$^{-1}$ for
India while in Germany the interval is restricted to $0.81$ ms$^{-1}$ 
to $ 0.96$ms$^{-1}$. For larger $N$ this effect is much more pronounced; for example
the density range for $N=34$ in India is $1.4 1\;$m$^{-1}$ to $ 2.6 1\;$m$^{-1}$ 
while in Germany it is $1.5 1\;$m$^{-1}$ to $ 2.0 1\;$m$^{-1}$. Thus the 
differences between individual crossing times and mean
distances to adjacent persons for a fixed $N$ is higher in India indicating a
more unordered character of the movement; even though the Indian composition of
test persons is more homogeneous than the German group. This is affirmed by
examination of the video recordings. Another difference is the rate of change 
of speed with change in density; this seems to be much higher for mid to high 
level densities (i.e., between $1$ to $2\;$m$^{-1}$) in Germany than in India.

\begin{figure}[htb]
\centering
    \includegraphics[width=0.45\columnwidth,angle=-90]{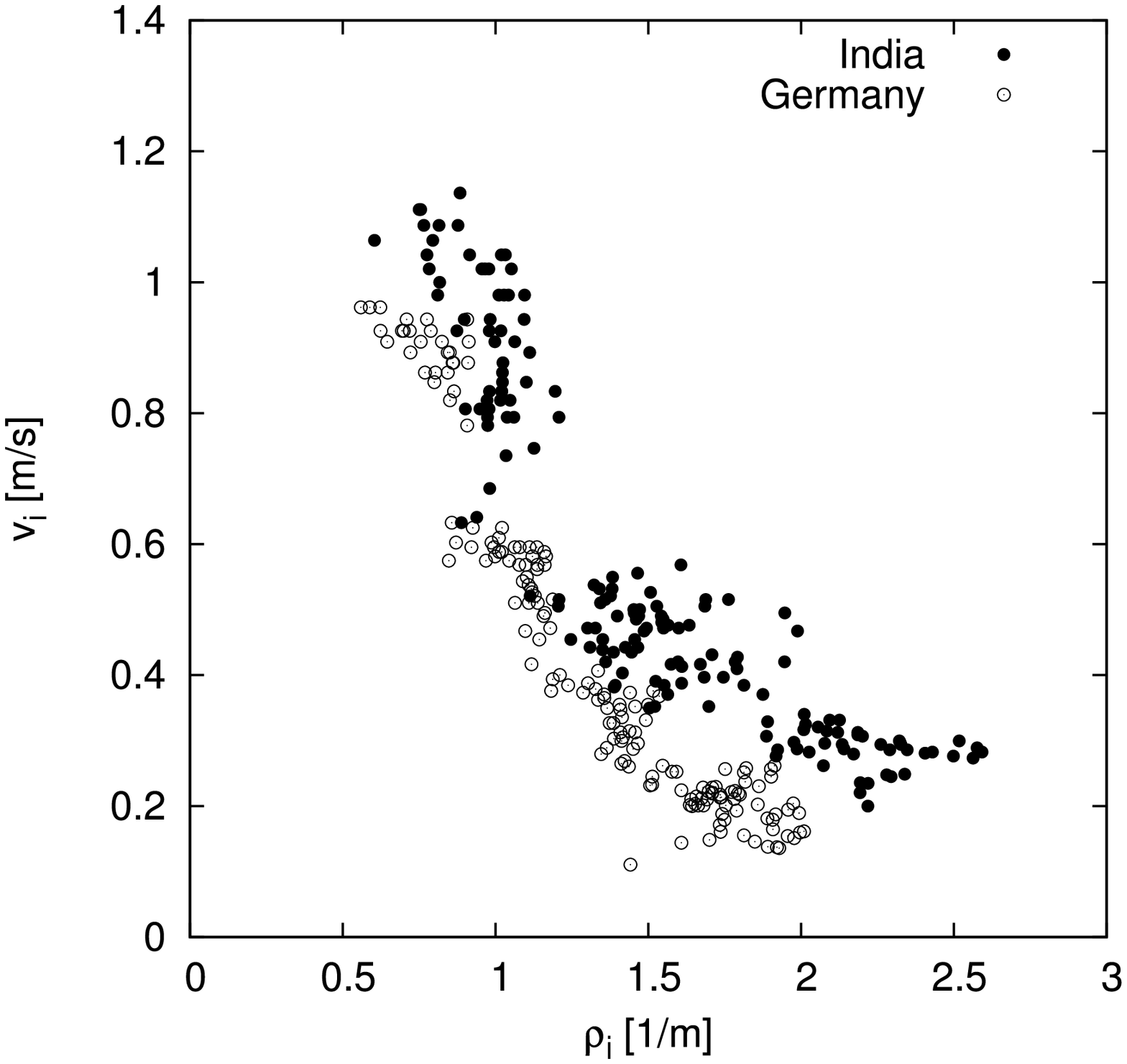}
\quad
    \includegraphics[width=0.45\columnwidth,angle=-90]{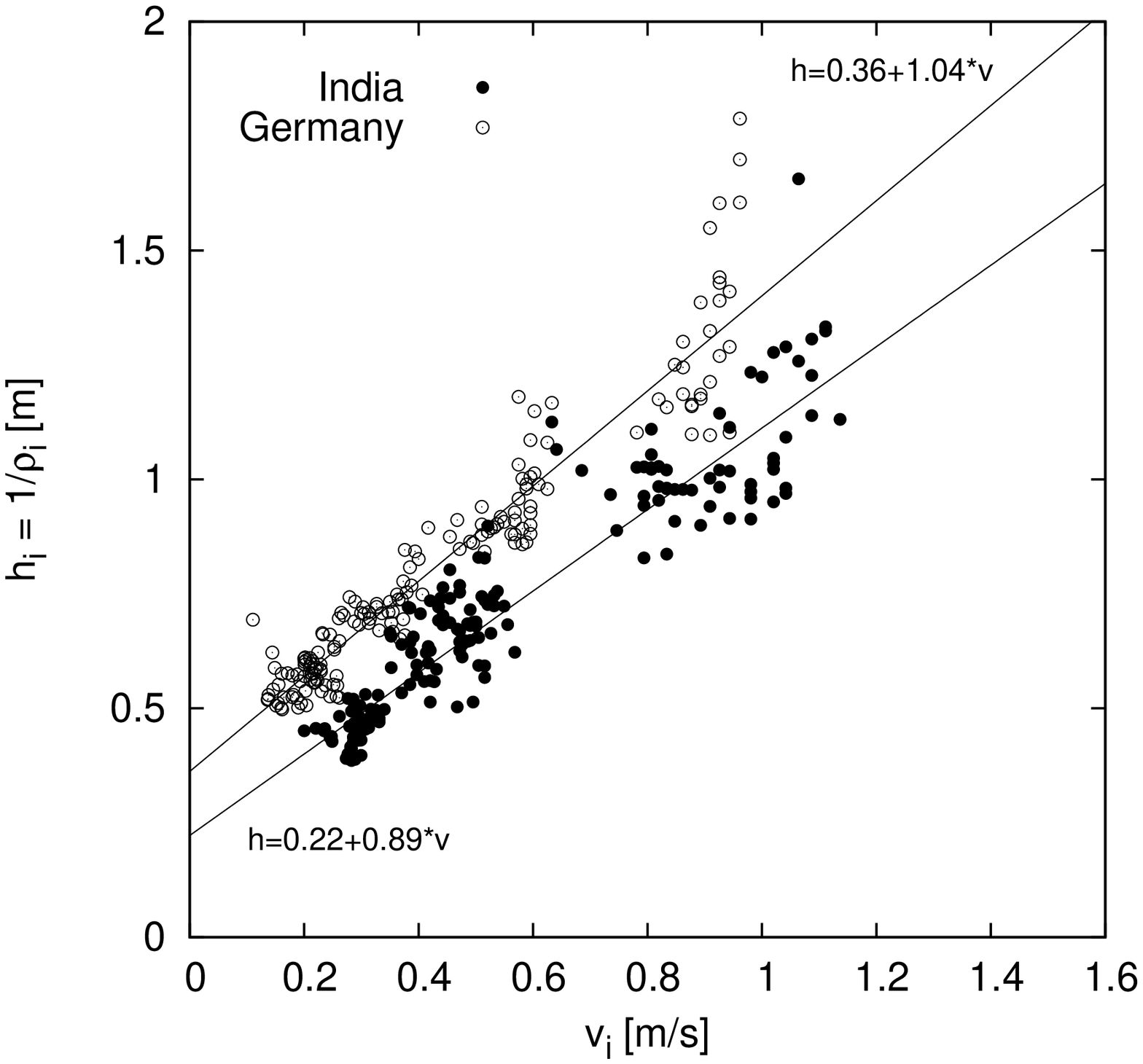}
\caption{{\bf Left: }Relation between speed ($v$) and density ($\rho$) from
  Indian and German study, respectively in closed condition. {\bf Right:}
  shows the fitted linear $h$-$v$ relations for the Indian and German data,
  respectively. Table~\ref{MEANHV} presents more details of the regressed
  lines. As can be seen from the $R^2$ - values the simple linear relation $h
  = a + b\;v$ seems to explain much of the variation observed in the $h$-$v$
  data.
} \label{RHO-V-DIFF}
\end{figure}

But, the most important point observed here is that the speed-density
relationship is not linear both for India and Germany, see Figure
\ref{RHO-V-ALL}.
On the other hand, the distance headway ($h$) versus speed ($v$) plot (Figure
\ref{RHO-V-DIFF}, right) is more suitable for description through a linear
relationship. In order to study this observations made so far more closely
first linear relationships are fitted to the Indian and German data on $h$
versus $v$ and their various hypotheses are tested on the estimated
coefficients.

\subsection{Speed-headway relation}
Relationships of the form $h$=$a$+$b\,v$ are fitted to the Indian and the
German data. The values of $a$ and $b$ for the Indian and the German data are
shown in Table~\ref{MEANHV} along with other statistics. The fitted lines are
also plotted with the data and are shown in Figure $6$ (right). From 
the $t$-statistic values and the $R^2$ values shown in Table~\ref{MEANHV} it 
can be said that obtained fits are statistically sound. Note that in the 
relation $h$=$a$+$b\,v$, $a$ represents the minimum personal space and $b$ 
denotes the sensitivity of $h$ to $v$, or $\frac{dh}{dv}$.

\subsubsection{Study on minimum personal space}

As shown in Table~\ref{MEANHV} the estimated minimum personal space $a$ for
India is $0.22$ m and that for Germany is $0.36$ m. The standard errors of $a$ 
is $S_{aI}=0.02$ for India and $S_{aG}=0.01$ for Germany. The hypothesis that 
minimum personal space obtained from India, $a^I$ and that from Germany $a^G$ are 
the same (i.e., Null hypothesis $H_0:\;a^G-a^I = 0$ and alternate 
hypothesis $H_1:\;a^G -a^I \neq 0$) was tested. Under the standard normality 
assumption and since the number of data points obtained from experiments was 
large the expression

\begin{equation}
  z = \frac{a^G-a^I}{\sqrt{S^2_{aG}+S^2_{aI}}}
\end{equation}

can be assumed to be normally distributed. $S_{aI}$ is the standard error for
the data from India and $S_{aG}$ for Germany. If calculated $z$ value is more
than the $z_{critical}$ value from table the Null hypothesis can be rejected,
otherwise not.

For the above values of $a^I$, $a^G$, $S_{aI}$ and $S_{aG}$, $z$ from the above
expression turns out to be $6.46$. Since, this is more than $1.96$, 
the $z_{critical}$ at $95\%$ level of confidence for a two tailed test the Null 
hypothesis can be rejected. In other words estimated minimum personal space 
observed in India and Germany are statistically different. Here it is more for 
Germany than for India. Thus, one should observe higher jam density in India 
than in Germany.

\subsubsection{Study on change of distance headway with speed}

The coefficient $b$ represents the rate of change of distance headway ($h$)
with speed ($v$). Alternatively this coefficient expresses how people react (in
terms of speed) when the space ahead of them becomes restricted. As shown in the
Table~\ref{MEANHV} $b^I = 0.89$ and $b^G = 1.04$. The standard errors of $b$ is
$S_{bI}=0.03$ for India and $S_{bG}=0.03$ for Germany. The Null hypothesis ($H_0$:
$b^G  - b^I = 0 $ and alternate hypothesis $H_1$: $b^G  - b^I \neq 0$) was
tested. The test proceeds along similar line, as expressed in the previous
section. The $z$-value in this case comes out to be $3.99$ and $z_{critical}$
at $95\%$ level of confidence for a two tailed test is the same as in the
previous case. Since $z > z_{critical}$, the Null hypothesis that there is no 
difference in $b$ for India
and Germany can be rejected. Since $b^G$ is more than $b^I$, for increase in speed
Germans require more space ahead than Indians. Thus one can say that Germans
are more sensitive to high density conditions and decrease speed with increasing
density more quicker than the Indians.

\begin{table}[h]
\begin{center}
\caption[dummy]{Statistical measures for $h-v$ relationship in closed corridor
over cultures.}
\label{MEANHV}

\vspace*{10pt}

\begin{tabular}{ c  c  c  c  c }
Data set &  Intercept & Slope   & $R^2$  & \# data \\
         &  $a [m]$   & $b [s]$ &        & points \\
\hline
\hline
India           & 0.22                 &  0.89                   & 0.89  & 139
\\
$L_{Cor}=17.3m$  & (t statistic = 12.76) &  (t statistic = 32.99)  &        & \\
\hline
Germany  & 0.36                 & 1.04                   & 0.91  & 170   \\
$L_{Cor}=17.3m$  & (t statistic = 28.02) &  (t statistic = 40.33)  &        & \\
\end{tabular}
\end{center}
\end{table}

\begin{table}[h]
\begin{center}
\caption[dummy]{Statistical measures for $h-v$ relationship in closed corridor
over lengths of corridor.}
\label{MEANL}

\vspace*{10pt}

\begin{tabular}{ c  c  c  c  c }
Data set &  Intercept & Slope   & $R^2$  & \# data \\
         &  $a [m]$   & $b [s]$ &        & points \\
\hline
\hline
India           & 0.22                 &  0.89                   & 0.89  & 139
\\
$L_{Cor}=17.3m$  & (t statistic = 12.76) &  (t statistic = 32.99)  &        &
\\
\hline
India           & 0.25                  &  0.88                  & 0.90  & 246
\\
$L_{Cor}=34.6m$  & (t statistic = 22.83) &  (t statistic = 47.53)  &        &
\\
\end{tabular}
\end{center}
\end{table}


\subsection{Influence of corridor length}

\begin{figure}[htb]
\centering
    \includegraphics[width=0.42\columnwidth,angle=-90]{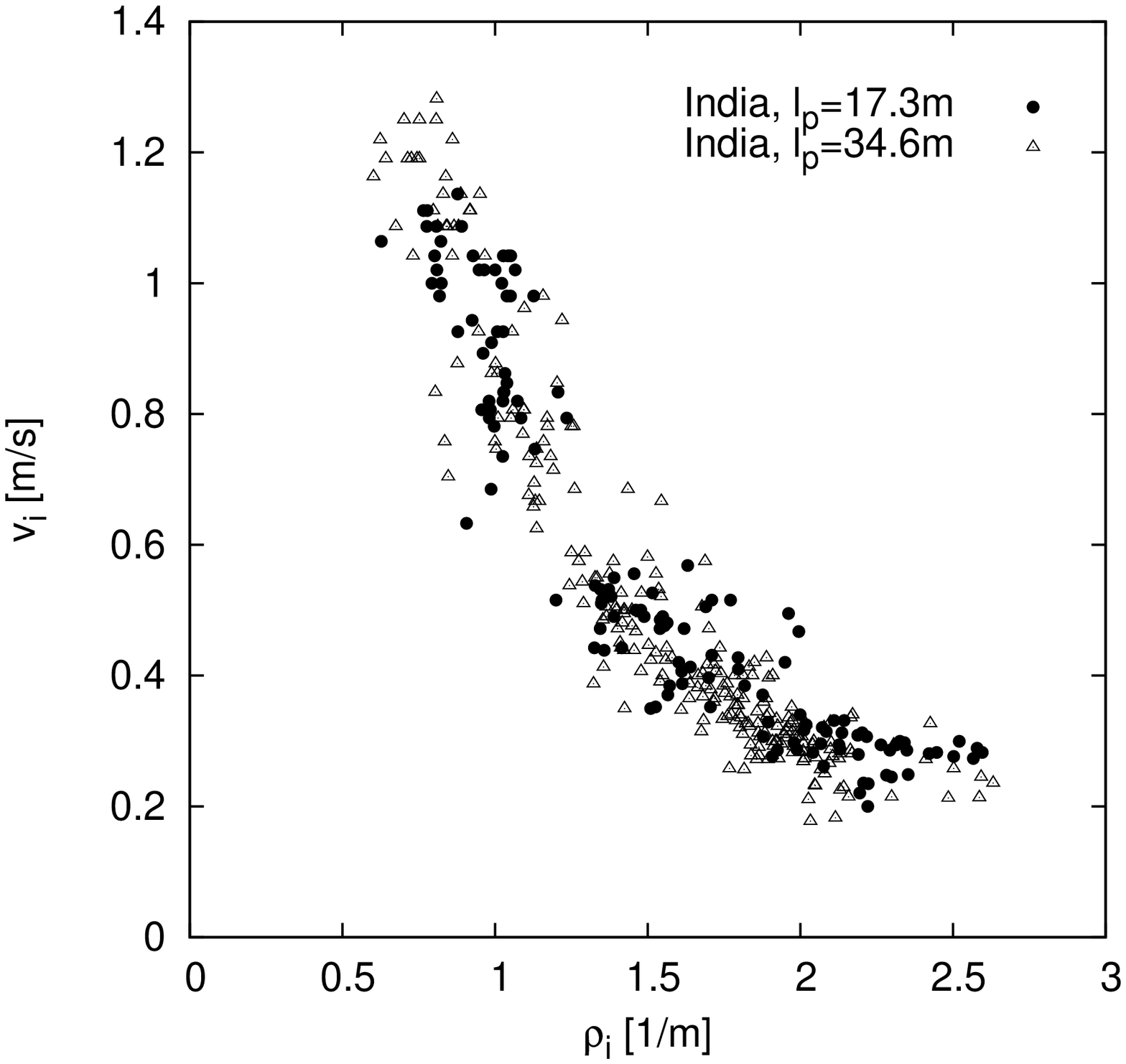}
\quad
    \includegraphics[width=0.42\columnwidth,angle=-90]{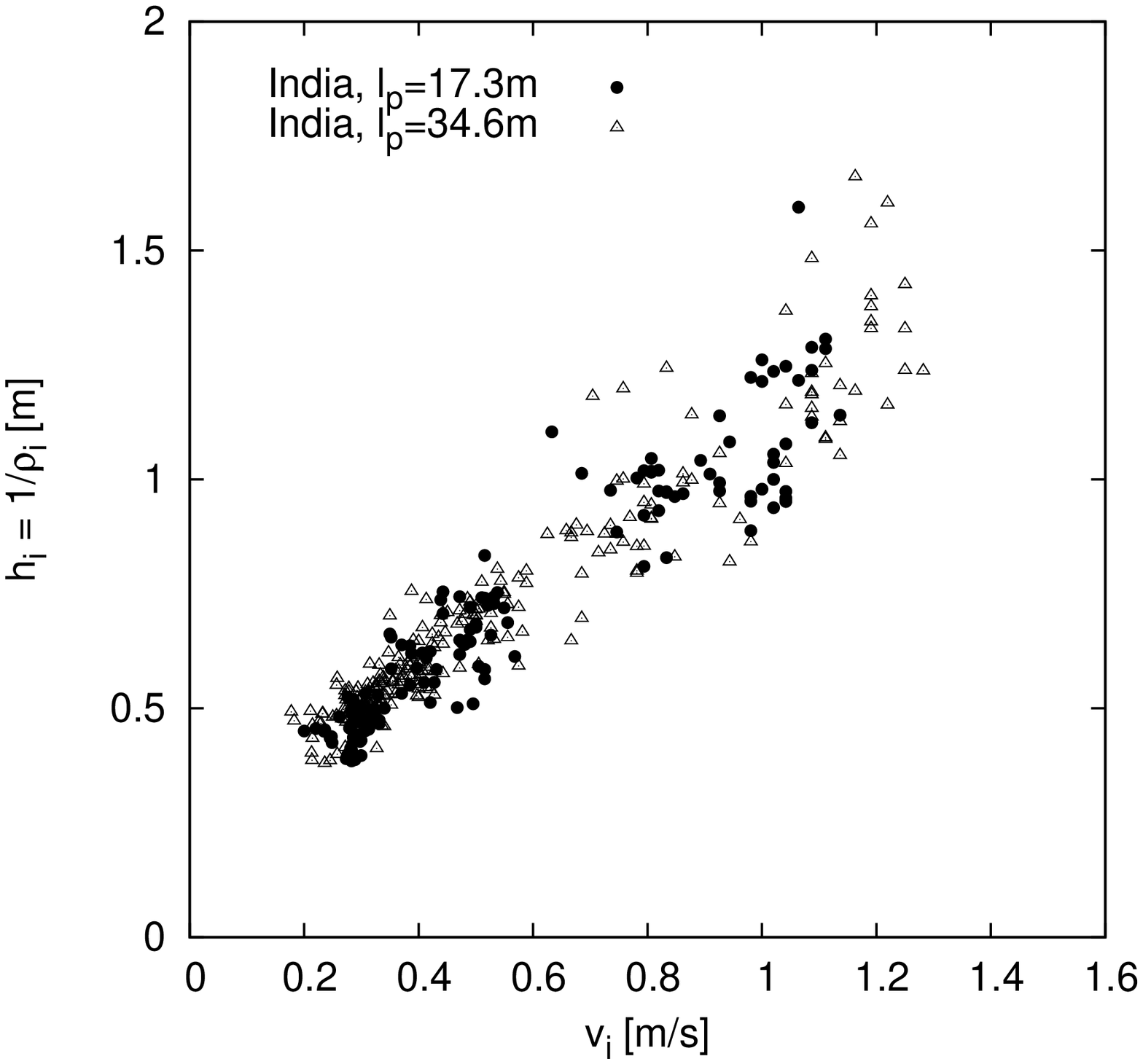}

\caption{Speed-density data from India (Left) and speed-headway data from India
(Right) for different lengths of corridor.
} \label{RHO-V-ALL-LENG}
\end{figure}

The distance headway ($h$)-speed ($v$) data in closed corridor for different
length of corridor $l_p=17.3$ m and $34.6$ m obtained from India are plotted in
Figure \ref{RHO-V-ALL-LENG}. To achieve comparable densities for the double
length of corridor
we choose the following numbers of test persons: $N=30, 40, 50, 60, 68$. The
values of $a$, $b$ and $R^2$ for a fitted relation $h$=$a$+$b\,v$ are as shown in 
Table~\ref{MEANL}. On conducting hypothesis tests on the difference of
intercepts and on the difference of slope terms for the small length and large length
section it was found that none of the differences are statistically
significant. That is, length of the corridor has no impact on the $h$-$v$
relation.

\section{Conclusions}

In this study fundamental diagrams from different cultures, namely German and
Indian is presented.
Fundamental diagrams for two different lengths of corridor is also presented.
Same measurement method was followed throughout and results were compared
statistically using $t$-test and $z$-test. The following are the observations:
The free flow speed is the same for the Indian and the German group of test
persons suggesting that when by themselves German and Indian pedestrians move
in a similar manner. Estimated minimum personal space for the German group is
more than that for Indian group indicating a higher jam density for India.
Increase in distance headway with increase in speed for the German group
is more than that of the Indian group indicating that Indians are less
sensitive to increase in density compared to the Germans. It is believed that
one cause for these differences is due to culture. For explanation of this
statement the term security distance has to be introduced.
The actual gap kept (i.e., the distance headway $h$) is composed of
the required space for movement and the gap one wants to keep with his
predecessor. This gap can be named as security distance. By visual inspection
it appears that the Indian group of test persons are less concerned about the
personal space of other persons and thus the security distance is smaller
compared to the German group. To verify this it requires a method by which the
dependence of fundamental diagram on security distance can be measured.
Measurement of security distance from experimental data is difficult
because the required space for movement changes temporarily due to the
movement in steps. To test this idea it is planned to
build a model where the security distance is a parameter. In addition for
strong conclusion about cultural differences more experiments with various
cultures and test group combinations are needed. Another important finding
is that the length of the corridor does not have an effect on the
fundamental diagram.\\

{\bf Acknowledgment}

The authors are thankful to IIT Kanpur Transportation Laboratory (India) and
J\"ulich Supercomputing Centre (Germany) for providing experimental
facilities. The project was partially funded by the Deutsche
Forschungsgemeinschaft (DFG) under Grant-Nr.: SE 1789/1-1.\\


\end{document}